\documentclass[letterpaper]{JHEP}
\usepackage{epsfig}
\usepackage{graphicx}
\usepackage{amssymb,amsmath}
\usepackage{epstopdf}

\def\be{\begin{eqnarray}}
\def\ee{\end{eqnarray}}

\title{Branes and massive IIA duals of 3d CFT's}

\author{Oren Bergman\\
Department of Physics\\
Technion, Israel Institute of Technology\\
Haifa 32000, Israel\\
\email{bergman@physics.technion.ac.il}}

\author{Gilad Lifschytz\\
Department of Mathematics and Physics and CCMSC \\
University of Haifa at Oranim\\
Tivon 36006, Israel \\
\email{giladl@research.haifa.ac.il}}

\date{}

\abstract{We describe brane configurations that interpolate
between the ${\cal N}=6$ $AdS_4\times \mathbb{C}P^3$ background of Type IIA
supergravity and the ${\cal N}=0$ $AdS_4\times \mathbb{C}P^3$ background of
massive Type IIA supergravity.
Using the T-dual Type IIB configurations we prove that this leads to unequal Chern-Simons
levels in the dual gauge theory, and find the precise relation between 
the parameters of the gauge theory and the RR fluxes of the background.
This provides further evidence for the conjecture of Gaiotto and Tomasiello
about the CFT dual of the massive Type IIA background.}

\begin{document}

\setcounter{equation}{0}
\setcounter{figure}{0}
\setcounter{table}{0}

\section{Introduction}

Massive Type IIA supergravity is unique in that it does not seem to fit
within the M-theory framework, even though it is a maximally supersymmetric theory.
It was originally formulated as a variant of Type IIA supergravity, in which the (NSNS) 2-form
eats the (RR) 1-form and becomes massive \cite{Romans}. This also gives a cosmological constant
$\Lambda\sim m^2$. The mass was subsequently interpreted as an RR 0-form field strength $F_0$,
associated with D8-branes in Type IIA string theory \cite{Polchinski}.
However it is not yet known whether and how D8-branes lift to M-theory.
The non-zero cosmological constant also makes it harder to find supersymmetric solutions.
For example, Freund-Rubin type solutions of the form $AdS_4\times M_6$,
where $M_6=S^6, \mathbb{C}P^3, S^4\times S^2$ or $S^2\times S^2\times S^2$, and with $F_6$ flux were 
found in \cite{Romans}, but they were all non-supersymmetric.
They were also shown later to be unstable \cite{DeWolfe}.
More recently a host of ${\cal N}=1$ supersymmetric $AdS_4\times M_6$ solutions,
with $M_6$ a nearly-Kahler space, have been found \cite{Cvetic,Tomasiello,Lust}. 
These solutions involve both $F_6$ flux and $F_2$ flux in the compact space.
A question that immediately arises is what are the dual 3d superconformal field theories?

For many similar backgrounds in {\em massless} Type IIA supergravity we know the answer.
The dual field theories are quiver Chern-Simons-Matter (CSM) theories,
with CS levels that sum to zero $\sum_i k_i = 0$.
In this case the connection is made through M2-branes, whose low energy dynamics
is described by these superconformal CSM theories.
At large $N$ these provide a large class of explicit
$AdS_4/CFT_3$ dual pairs with supersymmetries ranging from ${\cal N}=8$ down to ${\cal N}=1$.
The simplest case is the theory of $N$ M2-branes on the orbifold $\mathbb{C}^4/\mathbb{Z}_k$,
which is an ${\cal N}=6$ CS theory with gauge group $U(N)_k\times U(N)_{-k}$ 
and matter fields in the bi-fundamental representation \cite{ABJM}.
For $k=1$ and $k=2$ it describes M2-branes in flat space and $\mathbb{R}^8/\mathbb{Z}_2$, respectively,
and the supersymmetry is enhanced non-perturbatively to ${\cal N}=8$ \cite{Rey,Kwon}.
The field theory has an effective 'tHooft coulping given by $\lambda=N/k$.
At large $N$ and for $k\ll N^{1/5}$ the theory is dual to M theory on $AdS_4\times S^7/\mathbb{Z}_k$,
with $N$ units of $G_7$ flux on $S^7/\mathbb{Z}_k$, and
for $N^{1/5}\ll k \ll N$ the dual theory is Type IIA string theory
on $AdS_4\times \mathbb{C}P^3$, with $F_6 = N$ and $F_2 = k$ (we are using an abuse of notation
where $F_6$ and $F_2$ stand for the number of their flux quanta on $\mathbb{C}P^3$ and 
$\mathbb{C}P^1\subset\mathbb{C}P^3$, respectively).  
The duality can also be extended to unequal ranks, $U(N+l)_k \times U(N)_{-k}$ with $l\leq k$ \cite{ABJ}.
The M theory dual in this case has $[G_4]\in\mathbb{Z}_k$, 
and the Type IIA dual has a $B$ field
holonomy $l/k$. The latter also implies an additional RR flux $F_4 = l$, corresponding to 
$l$ D4-branes wrapped on the 2-cycle $\mathbb{C}P^1\subset \mathbb{C}P^3$.

In a recent development, Gaiotto and Tomasiello have proposed 
that the CFT duals of massive Type IIA $AdS_4\times M_6$ solutions are 
quiver CSM theories of the same type,
except that the sum of the CS levels is given by $\sum_i k_i = F_0$ \cite{GT}.
In particular, they studied deformations of the ${\cal N}=6$ theory to 
$U(N)_{k_1}\times U(N)_{k_2}$ with $k_1\neq {-k_2}$, that preserve ${\cal N}=0,1,2$ and 3 supersymmetry, and $SO(6), SO(5), SO(2)_R\times SO(4)$ and $SO(3)_R\times SO(3)$
global symmetry, respectively.
It was argued that in each case the theory flows to a unique CFT with
the corresponding supersymmetry and global symmetry.
They further conjectured that the ${\cal N}=0$ and ${\cal N}=1$ deformations are dual 
to $AdS_4\times \mathbb{C}P^3$ solutions with $F_0=k_1+k_2$,
with the $SO(6)$ invariant (Fubini-Study) metric on $\mathbb{C}P^3$ in the ${\cal N}=0$ case,
and the $SO(5)$ invariant (squashed) metric in the ${\cal N}=1$ case.\footnote{The solutions 
dual to the ${\cal N}=2$ and ${\cal N}=3$ deformations are not known
exactly, but have been constructed perturbatively for small $F_0$ \cite{GT2}. A different 
${\cal N}=2$ example
corresponding to a deformation of the $AdS_4\times M^{(1,1,1)}$ solution of M theory
was given in \cite{Petrini}.}
In the more general case of unequal ranks the field theory has four parameters
which are related to the fluxes of $F_6,F_4,F_2$ and $F_0$.
The main evidence for this conjecture
comes from considering the properties of D0-branes and D2-branes in the massive Type IIA backgrounds,
as compared with the original ${\cal N}=6$ solution.
In particular tadpole cancellation on the D0-brane requires $F_0$ strings to end on it, due to the
coupling $F_0 A$. 
This agrees with the fact that the dual di-monopole operators 
of the field theory have extra gauge indices when $k_1\neq -k_2$, which must be saturated
by $|k_1+k_2|$  semi-infinite Wilson lines.
The analogous coupling on the D2-brane held at a fixed radial position in $AdS_4$ gives a level $F_0$ 
CS term. In the field theory this corresponds to Higgsing the $U(1)_-$ in a $U(1)\times U(1)$ subgroup.
For $k_1=-k_2$ this produces a Maxwell action for the $U(1)_+$ \cite{Mukhi},
but for $k_1\neq -k_2$ there is also a remnant level $(k_1+k_2)$ CS term,
in agreement with the supergravity result.

\medskip

Our goal in this paper is to provide more evidence for this conjecture.
As an additional consistency check, consider a D4-brane probe at a fixed radial position,
and wrapping the 2-cycle 
$\mathbb{C}P^1\subset \mathbb{C}P^3$. This is the ``fractional" D2-brane \cite{ABJ}.
In the background with $k$ units of $F_2$ flux one gets a 3d CS term from the 5d RR coupling
on the D4-brane:
\be
 \int_{R^{1,2}\times \mathbb{C}P^1} C_1 \wedge F \wedge F =  
  \int_{R^{1,2}\times \mathbb{C}P^1} F_2 \wedge A \wedge F =
 k \int_{R^{1,2}} A\wedge F \,.
\ee
A full D2-brane is equivalent to a D4-anti-D4-brane pair, with one unit of worldvolume magnetic flux
through the $\mathbb{C}P^1$ on either the D4-brane or the anti-D4-brane.
For $F_0=0$ the D4-brane gets a level $k$ CS term, and the anti-D4-brane gets a level $-k$ CS term, 
due to the opposite sign of their RR couplings.
This is exactly as expected in the ${\cal N}=6$ theory.
If $F_0 = q \neq 0$, the worldvolume magnetic flux leads to
an additional CS term on either the D4-brane or the anti-D4-brane (but not both):
\be
\int_{R^{1,2}\times \mathbb{C}P^1} F_0 \, A\wedge F\wedge F = q \int_{R^{1,2}} A\wedge F \,. 
\ee
So we get either $U(1)_{k+q}\times U(1)_{-k}$ or $U(1)_k\times U(1)_{-k+q}$.
This raises a puzzle, since it seems to imply that there are two distinct field theory duals
to a single solution with $F_0 = q$ and $F_2 = k$ (again with an abuse of notation):
$U(N_1)_{k+q}\times U(N_2)_{-k}$ or $U(N_1)_k\times U(N_2)_{-k+q}$.
In fact a more general question is how precisely a non-vanishing $F_0$ affects
each of the two CS levels. In principle the condition $k_1 + k_2 =F_0$ can be satisfied 
in many ways. We will show that this question has a definite answer.

In what follows we will describe string theory backgrounds that interpolate between
the massless Type IIA solution with ${\cal N}=6$ and the massive Type IIA solution
with ${\cal N}=0$.
In particular in section 2 we will study a D8-brane deformation of the massless IIA solution,
and in section 3 we will study its T-dual realization in a Type IIB brane configuration.
The latter will turn out to be a simple generalization of the brane configuration used in \cite{ABJM}, 
and we will show explicitly how it leads to different CS levels for the two gauge groups.
With the aid of the Type IIB description, we will then determine the precise relation between the RR fluxes
and the field theory parameters in section 4, generalizing the result of \cite{AHHO} for the massless IIA solutions.
The appendix contains some relevant details on the geometry of $\mathbb{C}P^3$.

\section{D8-brane deformation}

We will concentrate on the ${\cal N}=0$ solution with the $SO(6)$ invariant metric on $\mathbb{C}P^3$.
This is the simplest and most symmetric deformation of the ${\cal N}=6$ solution.
Being non-supersymmetric, one might question the stability of this solution.
However for $F_0\ll F_2$ the solution is at least perturbatively stable.
This is because the most tachyonic state of the ${\cal N}=6$ solution with $F_0=0$ has 
$m^2=-2R_{AdS}^{-2}$,
which is well above the BF bound for $AdS_4$ of $-(9/4) R_{AdS}^{-2}$.
We expect an instability to arise as $F_0$ increases, since in
the opposite limit $F_0\gg F_2$ we begin to approximate the solution with $F_2=0$ of \cite{Romans},
which is unstable.

\begin{figure}[htbp]
\begin{center}
\epsfig{file=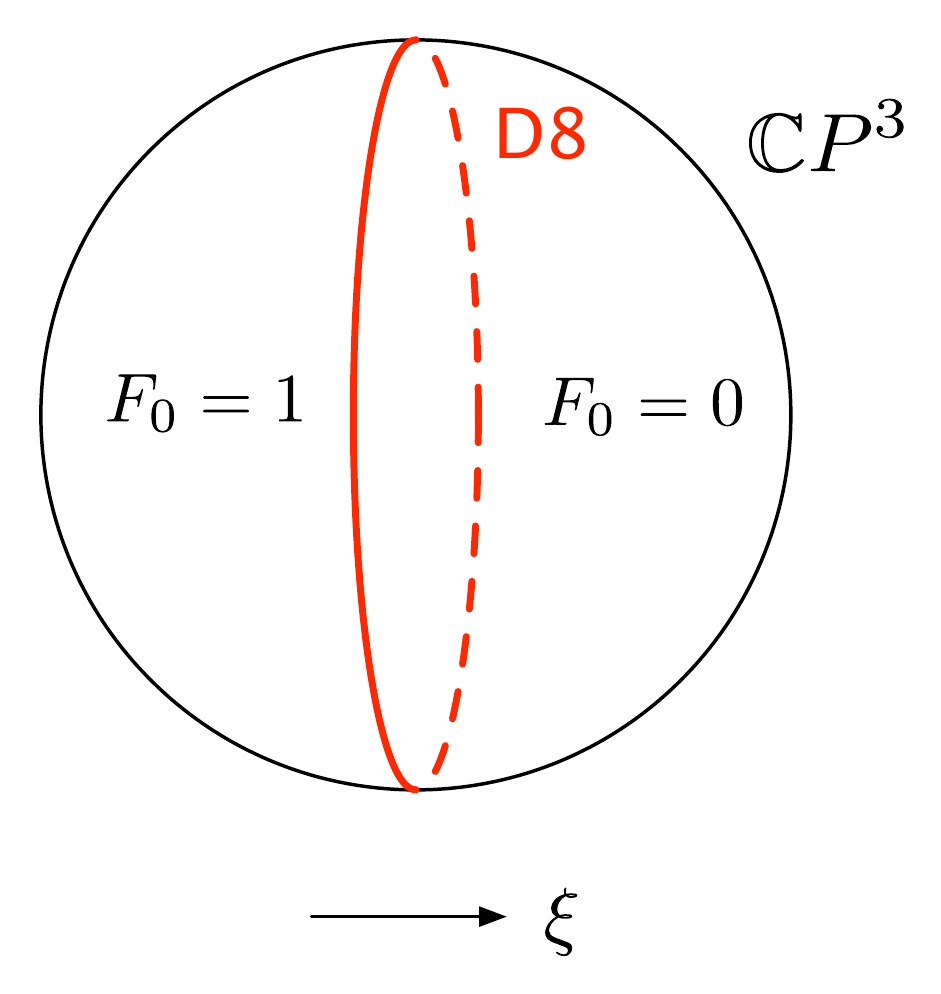,scale=0.43}
\hskip 1cm
\epsfig{file=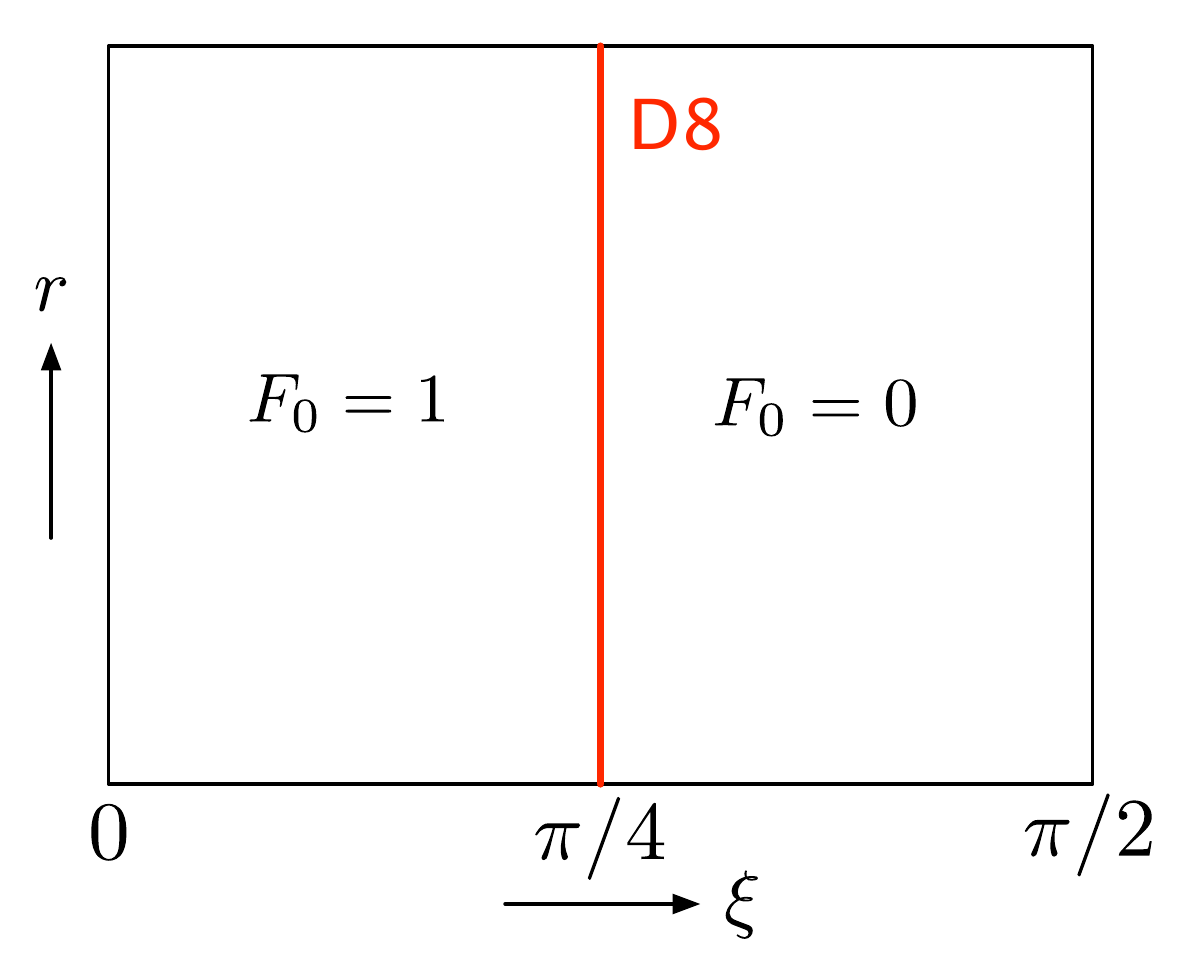,scale=0.43}
\caption{{\bf Two views of the D8-brane embedding in $AdS_4\times \mathbb{C}P^3$.}}
\label{M5}
\end{center}
\end{figure}

A D8-brane may be embedded into $AdS_4\times \mathbb{C}P^3$ by spanning
$AdS_4$ and a 5-dimensional subspace of $\mathbb{C}P^3$ defined by $\xi(r)$
(see the Appendix for details on the geometry of $\mathbb{C}P^3$).
This forms a domain wall  in $\mathbb{C}P^3$ across which the value of $F_0$ 
jumps by one unit (Fig.~\ref{M5}).
The induced metric on the D8-brane is given by
\be
R_s^{-2} ds^2_{D8} & = & {r^2\over 4} \left(-dt^2 + dx_1^2 + dx_2^2 \right)
+{1\over 4r^2}\left(1 + 4r^2(\xi'(r))^2\right) dr^2 \nonumber \\
&+& \cos^2\xi \sin^2\xi \left(d\psi+{\cos\theta_1\over 2}d\phi_1-{\cos\theta_2\over 2}d\phi_2\right)^2 \\
&+& {\cos^2\xi\over 4}\left(d\theta_1^2+\sin^2\theta_1d\phi_1^2\right)
+ {\sin^2\xi\over 4}\left(d\theta_2^2+\sin^2\theta_2d\phi_2^2\right)\,, \nonumber
\ee 
and the D8-brane action then has the form
\be 
S_{D8} \propto \int dr \, r^2 \sin^3(2\xi(r)) \sqrt{1+4r^2(\xi'(r))^2} \,.
\ee
However there are no stable embeddings.
There are constant solutions with $\xi =0$, $\pi/2$ and $\pi/4$.
The first two correspond to a shrunk D8-brane at the two ``poles" of the $\mathbb{C}P^3$,
and the third to an ``equatorial" embedding. The ``equator" in this case is a 
$T^{1,1}\subset\mathbb{C}P^3$.
It is easily seen that this embedding is unstable by considering the large $r$
asymptotic behavior of the general solution
\be
\xi(r) \approx {\pi\over 4} + cr^{-\Delta} \,.
\ee
The equation of motion gives a complex exponent $\Delta = 3/2 \pm i\sqrt{3}/2$,
or equivalently a tachyonic mass $m^2 = -3 R_{AdS}^{-2}$
that violates the BF bound.
Note that $R_{AdS}^2 = R_s^2/4$ in this background \cite{ABJM}.
The $T^{1,1}$ embedding is therefore unstable to ``slipping" towards one of the poles at $\xi=0$ or $\xi=\pi/2$.

Although this configuration is unstable, we can use it to construct a continuous, but not flat,
deformation from the $F_0=0$ solution to an $F_0 \neq 0$ solution:
start with $F_0$ zero size D8-branes at $\xi=0$, and pull them across $\mathbb{C}P^3$ to $\xi=\pi/2$.
Assuming we can use the probe approximation, the metric is unchanged, 
and the resulting massive Type IIA background 
is the ${\cal N}=0$ $AdS_4\times \mathbb{C}P^3$ solution.
Of course there are well known issues with treating D8-branes as probes,
stemming from the lack of decay with distance of their backreaction.
However we believe that the backreaction is under control in this case,
and that the D8-brane motion does indeed give a continuous deformation
from the $SO(6)$ symmetric massless Type IIA solution to the $SO(6)$ symmetric
massive Type IIA solution, although at intermediate stages $SO(6)$ is broken.
The situation is somewhat similar to the D8-anti-D8 configuration of the Sakai-Sugimoto model
\cite{SS}, in that we effectively have a D8-anti-D8 pair with a compact transverse direction.
For the Sakai-Sugimoto configuration the backreaction was studied in \cite{Cobi},
where it was shown to give a small correction to the background.


\begin{figure}[htbp]
\begin{center}
\epsfig{file=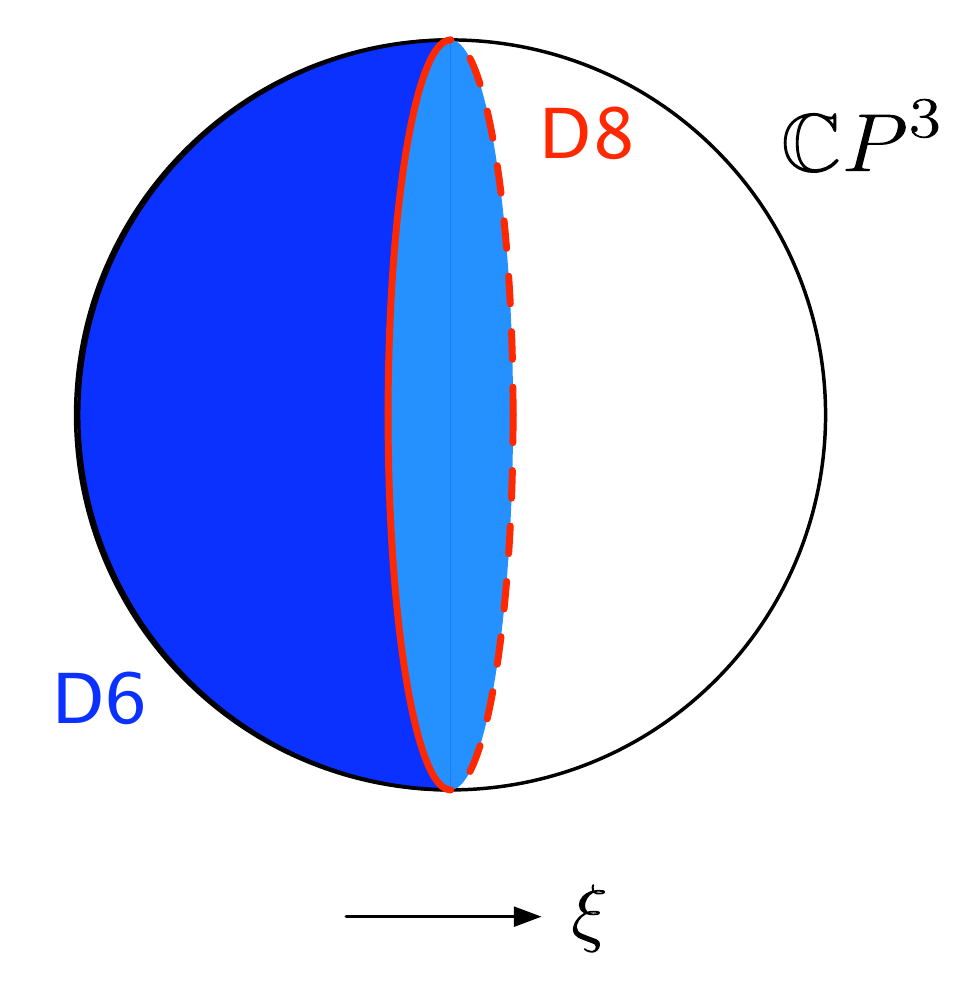,scale=0.43}
\hskip 1cm
\epsfig{file=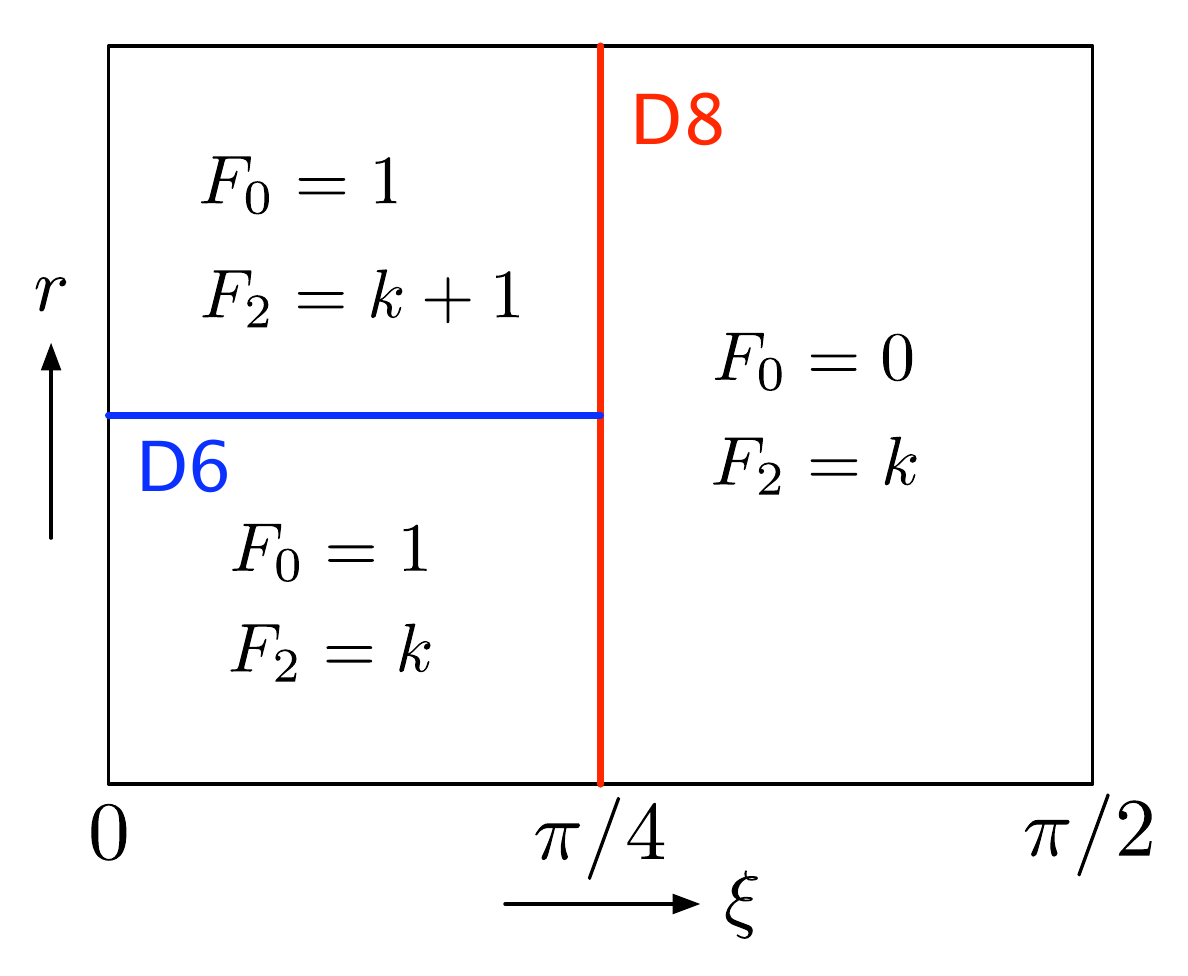,scale=0.43}
\caption{{\bf D8-D6 configuration in $AdS_4\times \mathbb{C}P^3$.}}
\label{D6brane}
\end{center}
\end{figure}

As a variant of the D8-brane deformation, we can also include D6-branes that end on the D8-brane.
The D6-branes are extended in the directions $t,x_1,x_2,\xi,\psi,\theta_1,\phi_1$,
and have a boundary at the location of the D8-brane in $\xi$ (Fig.~\ref{D6brane}).
We can describe their worldvolume as $R^{1,2}\times M_4$, where $M_4$ is a piece of 
the 4-cycle $\mathbb{C}P^2\subset\mathbb{C}P^3$, whose boundary is the 3-cycle $S^3\subset T^{1,1}$.
As the D8-brane is pulled across the $\mathbb{C}P^3$ the D6-brane covers the entire $\mathbb{C}P^2$.
The D6-brane itself forms a radial domain wall
across which the flux of $F_2$ on $\mathbb{C}P^1\subset\mathbb{C}P^3$ jumps by one unit.
It eventually falls into the horizon, and we end up with a massive Type IIA background with $F_2=k+1$.

\section{Type IIB brane configuration}

So far we have constructed a continuous interpolation, using branes, 
from the ${\cal N}=6$ $AdS_4\times \mathbb{C}P^3$ solution of massless IIA to 
the ${\cal N}=0$ $AdS_4\times \mathbb{C}P^3$ solution of massive IIA.
In this section we will move to Type IIB string theory, and describe the T-dual brane
configuration. We will give an independent derivation of the CS levels in the field theory, 
and we will show that the different CS levels are uniquely determined in terms of the brane configuration.
In particular, this will resolve the ambiguity of how $F_0$ contributes to the CS levels.

\begin{figure}[htbp]
\begin{center}
\epsfig{file=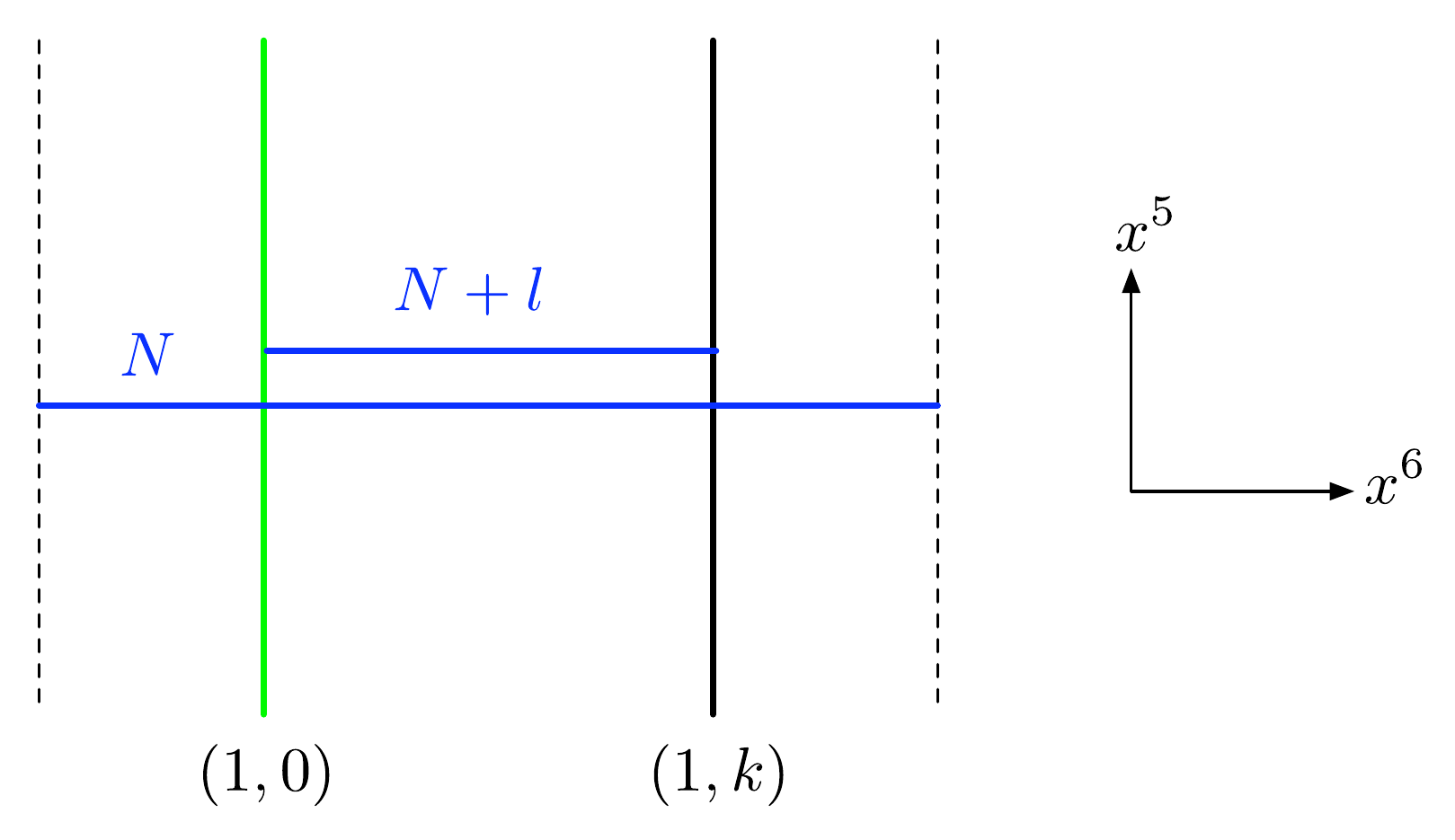,scale=0.4}
\caption{{\bf The Type IIB brane configuration for $U(N+l)_k\times U(N)_{-k}.$
The dashed lines are identified. We denote the NS5-brane as a $(1,0)$5-brane.}}
\label{ABJ_configuration}
\end{center}
\end{figure}

The Type IIB brane configuration used in the ABJM model
consists of an NS5-brane, a $(1,k)$5-brane (a bound state of an NS5-brane
and $k$ D5-branes), and a number of D3-branes arranged as follows \cite{KOO}: 
\begin{center}
\begin{tabular}{ccccccccccc}
& 0 & 1 & 2 & 3 & 4 & 5 & 6 & 7 & 8 & 9 \\
NS5 & $\bullet$ & $\bullet$ & $\bullet$ & $\bullet$ & $\bullet$ & $\bullet$ & & &  \\
$(1,k)5$ & $\bullet$ & $\bullet$ & $\bullet$ & $\cos\theta$ & $\cos\theta$ &$\cos\theta$ &
& $\sin\theta$ & $\sin\theta$ & $\sin\theta$ \\
D3 & $\bullet$ & $\bullet$ & $\bullet$ & & & & $\bullet$ & & & 
\end{tabular}
\end{center}
where the angle $\theta$ is the relative orientation of the two 5-branes in the
3-7, 4-8 and 5-9 planes, and is related to $k$ as $\tan\theta = k$ (for $g_s=1$ and $C_0=0$).
The coordinate $x^6$ is compact, and the D3-branes can either wind around it, or be suspended
between the two 5-branes (Fig.~\ref{ABJ_configuration}).
This describes a three-dimensional ${\cal N}=3$ gauge theory with a gauge group
$U(N+l)_k\times U(N)_{-k}$, and two bi-fundamental hyper-multiplets.
The two ranks $N+l, N$ are given by the number of D3-branes on either side of the circle
separated by the two 5-branes. For $l\leq k$ this theory flows 
in the IR to the ${\cal N}=6$ superconformal CSM theory with the same gauge group.
For $l>k$ the brane configuration appears to break supersymmetry due to the so-called
``s-rule" \cite{HW}, which suggests that supersymmetry is also broken in the gauge theory.
This is true in particular when $N=0$ and $l>k$ \cite{BHKK}, and explains 
supersymmetry breaking in the corresponding gauge theory \cite{Witten}.
However, as pointed out in \cite{AHHO,Evslin}, if $N\neq 0$
the supersymmetry bound on $l$ is actually higher, since in some cases
one can combine some of the wrapped D3-branes with some of the open D3-branes to
make multiply wrapped open D3-branes which satisfy the ``s-rule".
In fact these configurations can be obtained from a configuration with $l\leq k$ by
moving the NS5-brane around the circle multiple times. 
This suggests that in these cases supersymmetry is unbroken in the gauge theory,
and that it flows to one of the ${\cal N}=6$ superconformal CSM theories with $l\leq k$
via a 3d analog of the duality cascade \cite{AHHO,Evslin}.

The Type IIB brane configuration is related to the Type IIA $AdS_4\times \mathbb{C}P^3$ background 
by T-duality followed by a large $N$ and $k$ limit.
In \cite{ABJM} this was shown by first lifting the T-dual brane configuration to M theory,
taking the large $N$ limit, and then reducing back to Type IIA string theory.
But for our present purpose it is more useful to think about the direct route
(although it has not been worked out in detail)\footnote{The background dual to just the 5-branes
should be locally a cone over $\mathbb{C}P^3$.}, since we do not know
how to lift the D8-brane to M theory.
The object dual to the D8-brane on $AdS_4\times T^{1,1}$ is a D7-brane oriented as follows:
\begin{center}
\begin{tabular}{ccccccccccc}
& 0 & 1 & 2 & 3 & 4 & 5 & 6 & 7 & 8 & 9 \\
D7 & $\bullet$ & $\bullet$ & $\bullet$ & $\bullet$ & $\bullet$ & &  & $\bullet$ & $\bullet$ & $\bullet$
\end{tabular}
\end{center}
The D7-brane is pointlike in the 5-6 plane, and the the D3-D7 system has $ND=6$ 
directions with mixed boundary conditions.
This configuration breaks supersymmetry, and the D7-brane is repelled from the D3-branes
in the $x^5$ direction.
We can verify that this is the correct configuration by going through the transformation of the 
coordinates under the process
relating the Type IIB configuration to the Type IIA background.
This transformation is given explicitly by \cite{Takayanagi}:
\be
x^6 & = & \psi \nonumber \\
\vec{x}'_1 &=& \vec{x}_1 = r^2\cos^2\xi\left(\cos\theta_1,\sin\theta_1\cos\phi_1,\sin\theta_1\sin\phi_1\right) \nonumber\\
\vec{x}'_2 &=& \vec{x}_1 + k\vec{x}_2 = r^2\sin^2\xi\left(\cos\theta_2,\sin\theta_2\cos\phi_2,\sin\theta_2\sin\phi_2\right)\,,
\ee
where $\vec{x_1}\equiv (x^7,x^8,x^9)$ and $\vec{x}_2\equiv (x^3,x^4,x^5)$.
The metric in the primed coordinates is given by \cite{ABJM}
\be
ds^2 = {d\vec{x}_1' \cdot d\vec{x}_1'\over 2|\vec{x}_1'|} + {d\vec{x}_2' \cdot d\vec{x}_2'\over 2|\vec{x}_2'|} \,.
\ee
Let us verify that translating the D7-brane relative to the D3-branes in the $x^5$ direction 
corresponds to translating the D8-brane along $\xi$.
We need to compute the length of a path of fixed $\vec{x}_1$, where $\vec{x}_2$ varies
from $(0,0,0)$ to $(0,0,x^5)$. This is given by
\be 
L_5 = \int_{\vec{x}_1}^{\vec{x}_1+k(0,0,x^5)} {d|\vec{x}_2'|\over \sqrt{2|\vec{x}_2'|}}
= 2r\sin\left(\xi-{\pi\over 4}\right) \,.
\ee
The equatorial D8-brane embedding $\xi=\pi/4$ corresponds to the D7-brane intersecting the D3-branes,
which is an unstable configuration due to the repulsive force.
The D7-brane ``runs away" to $x^5\rightarrow \pm\infty$, in agreement with the ``slipping" instability
of the D8-brane.

\begin{figure}[htbp]
\begin{center}
\epsfig{file=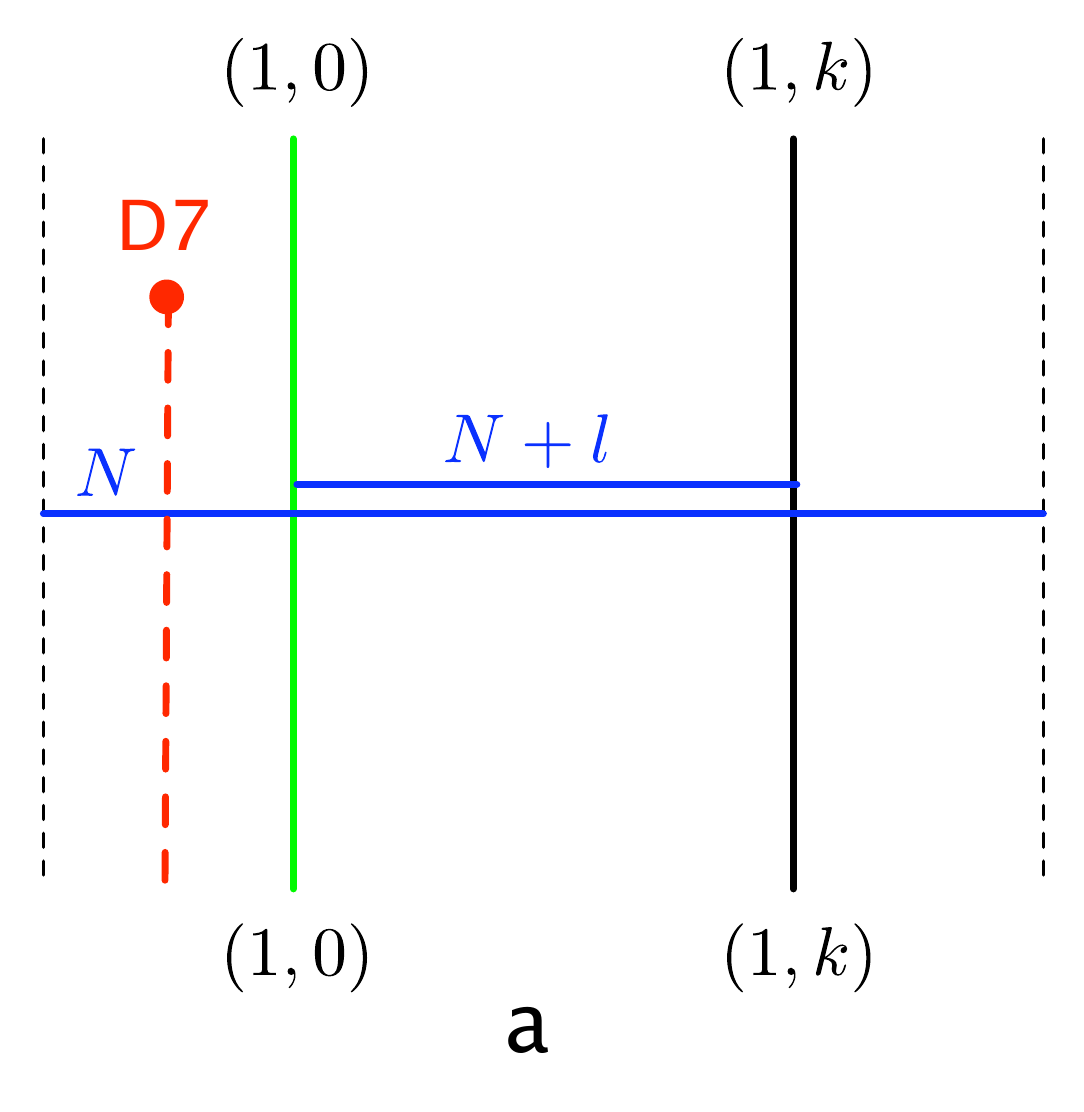,scale=0.4}
\hskip 1cm
\epsfig{file=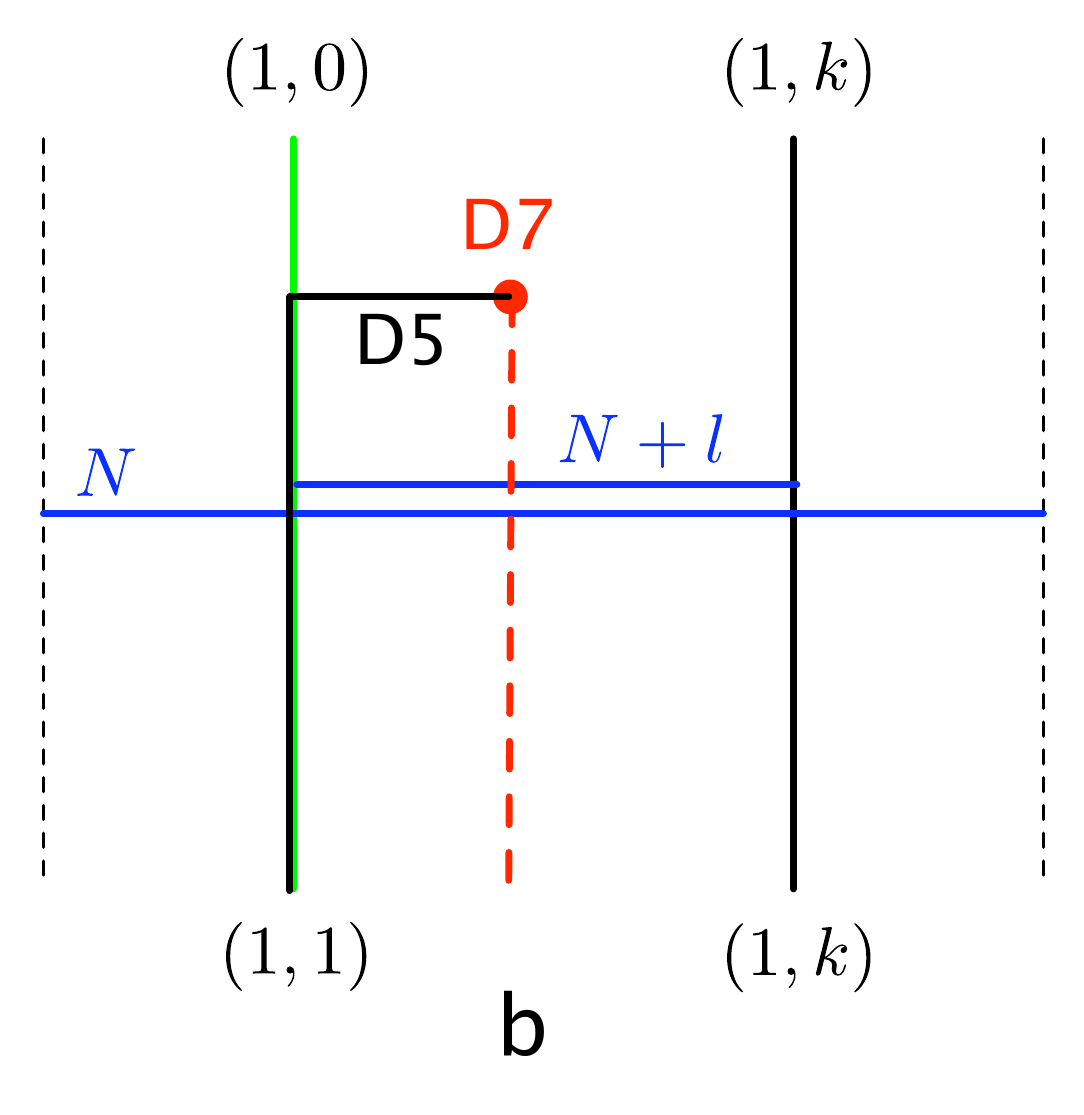,scale=0.4}
\caption{{\bf Type IIB brane configurations for $U(N+l)_k\times U(N)_{-k+1}$.}}
\label{D7_deformation}
\end{center}
\end{figure}

The deformation described in the previous section therefore corresponds to moving the D7-brane
from $x^5\rightarrow -\infty$ to $x^5\rightarrow +\infty$ across the D3-branes (Fig.~\ref{D7_deformation}a).
The key property of the D7-brane is that it sources
a monodromy for the RR scalar potential in the 5-6 plane 
$C_0\rightarrow C_0 + 2\pi$, which one can regard as occuring 
across a branch cut emanating from the D7-brane \cite{GSVY}.\footnote{The D7-brane is a special case
of a $(p,q)$7-brane with $p=1, q=0$ (a $(p,q)$7-brane is defined as the object on which a $(p,q)$ string
can end). The $(p,q)$7-brane sources an $SL(2,\mathbb{Z})$ monodromy for the dilaton-axion
$\tau \equiv C_0/(2\pi) + i\exp\Phi$ given by
\begin{equation*}
\tau \rightarrow {a\tau + b\over c\tau + d} \;\; , \;\; 
\left(
\begin{array}{cc}
a & b \\
c & d
\end{array}
\right) =
\left(
\begin{array}{cc}
1 - pq & p^2 \\
- q^2 & 1+pq
\end{array}
\right) \,.
\end{equation*}
}
We are of course free to choose the direction of the cut, which
we will take to be along $x^5$, {\em i.e.} parallel to the 5-branes,
from $x^5\rightarrow -\infty$ to the position of the D7-brane.
We will also comment later on the other choice, where the cut intersects the 5-branes.
As the D7-brane is taken to $x^5\rightarrow\infty$ we are left with a piecewise constant $C_0$ background
that jumps by $2\pi$ across the cut. This leads to an additional 3d CS term on the D3-brane that the cut intersects,
by integrating the 4d RR coupling:
\be
\label{RR_CS}
 \int_{R^{1,2}}\int_{x^6} C_0 \mbox{Tr} \left(F\wedge F\right) 
 =  2\pi S_{CS}\,.
\ee
The resulting gauge theory will then have either
$U(N+l)_k\times U(N)_{-k+1}$ or $U(N+l)_{k+1}\times U(N)_{-k}$,
depending on whether the D7-brane is on one side of the circle or the other.

These two theories are distinct, but are simply related by shifting $k\rightarrow k \pm 1$.
This is associated with the fact that the D7-brane and NS5-brane are linked in the sense of \cite{HW}.
Moving the D7-brane in the $x^6$ direction across the NS5-brane leads to the creation
of a D5-brane between them. This changes the part of the NS5-brane below 
the $x^5$ position of the D7-brane to a $(1,1)$5-brane (Fig.~\ref{D7_deformation}b),
and therefore shifts $k\rightarrow k-1$ (as can be seen by performing the $SL(2,\mathbb{Z})$
transformation $C_0\rightarrow C_0 - 2\pi$).
The resulting theory has $U(N+l)_{(k-1)+1}\times U(N)_{-(k-1)}$, and is identical
to the original theory.
From the low energy point of view we can therefore regard the configuration with the D7-brane
on one side and the configuration with the D7-brane on the other side, but with an additional (properly
oriented) D5-brane between the D7-brane and the NS5-brane, as identical.
We can therefore put all the D7-branes on one side, and describe the different configurations 
by the number of D5-branes connecting the D7-branes to the NS5-brane
(these violate the ``s-rule" if the number of D5-branes is greater than one, but supersymmetry
is anyway broken by the D7-brane).
In the Type IIA picture these configurations correspond precisely to the D8-brane embeddings
with additional D6-branes ending on them.
We will fix our convention by identifying the D8-brane embedding without attached D6-branes
with the D7-brane to the {\em left} of the NS5-brane without attached D5-branes.
This fixes the ambiguity mentioned in the introduction, {\em i.e.} the background with $F_0=q$
and $F_2=k$ is dual to the gauge theory 
with $U(N_1)_{k} \times U(N_2)_{-k+q}$.
Then the background dual to $U(N_1)_{k+q}\times U(N_2)_{-k}$, for example,
has $F_0=q$ and $F_2=k+q$.

\subsection{Alternative cuts and an alternative derivation of CS terms}

\begin{figure}[htbp]
\begin{center}
\epsfig{file=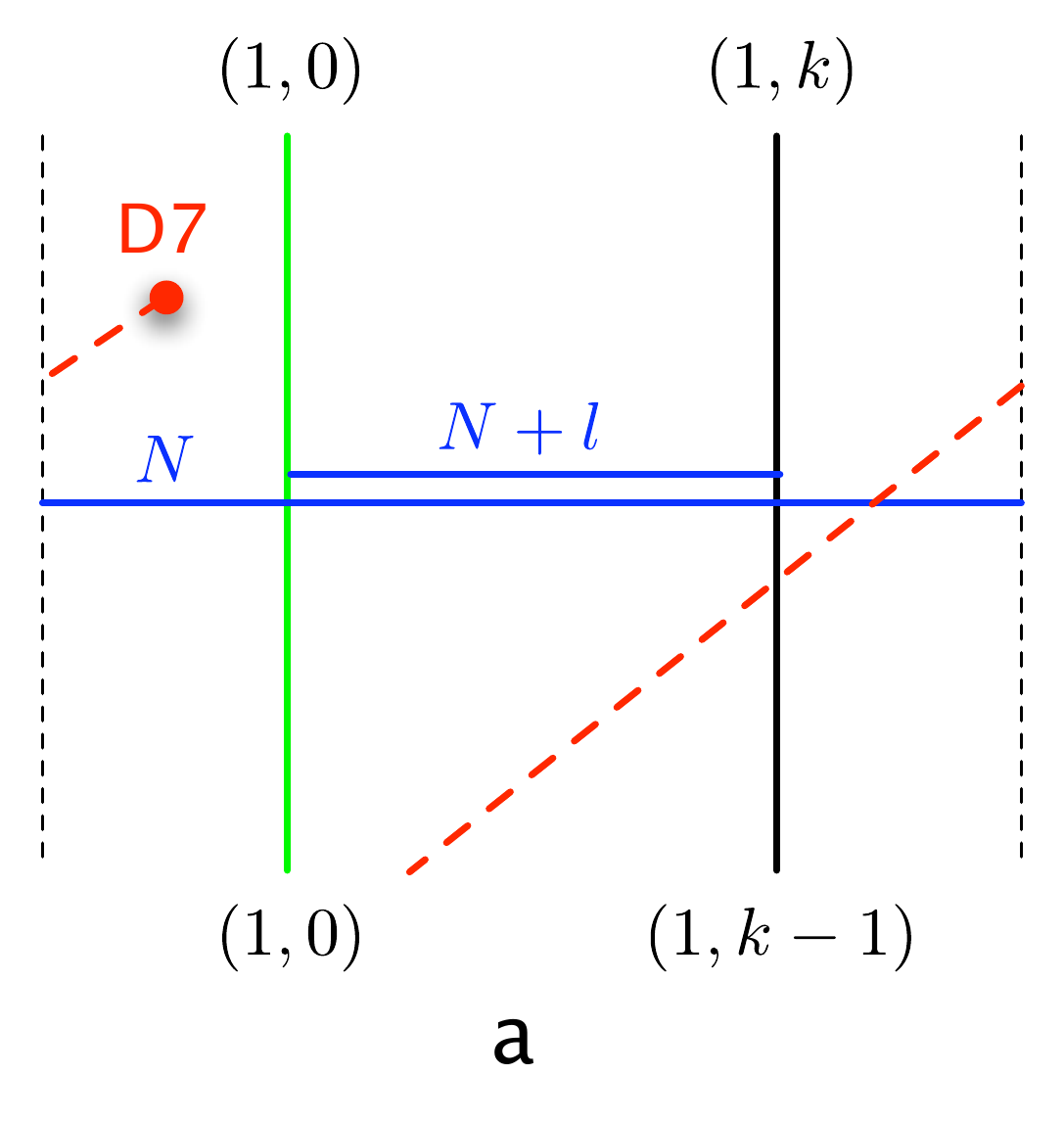,scale=0.4}
\hskip 1cm
\epsfig{file=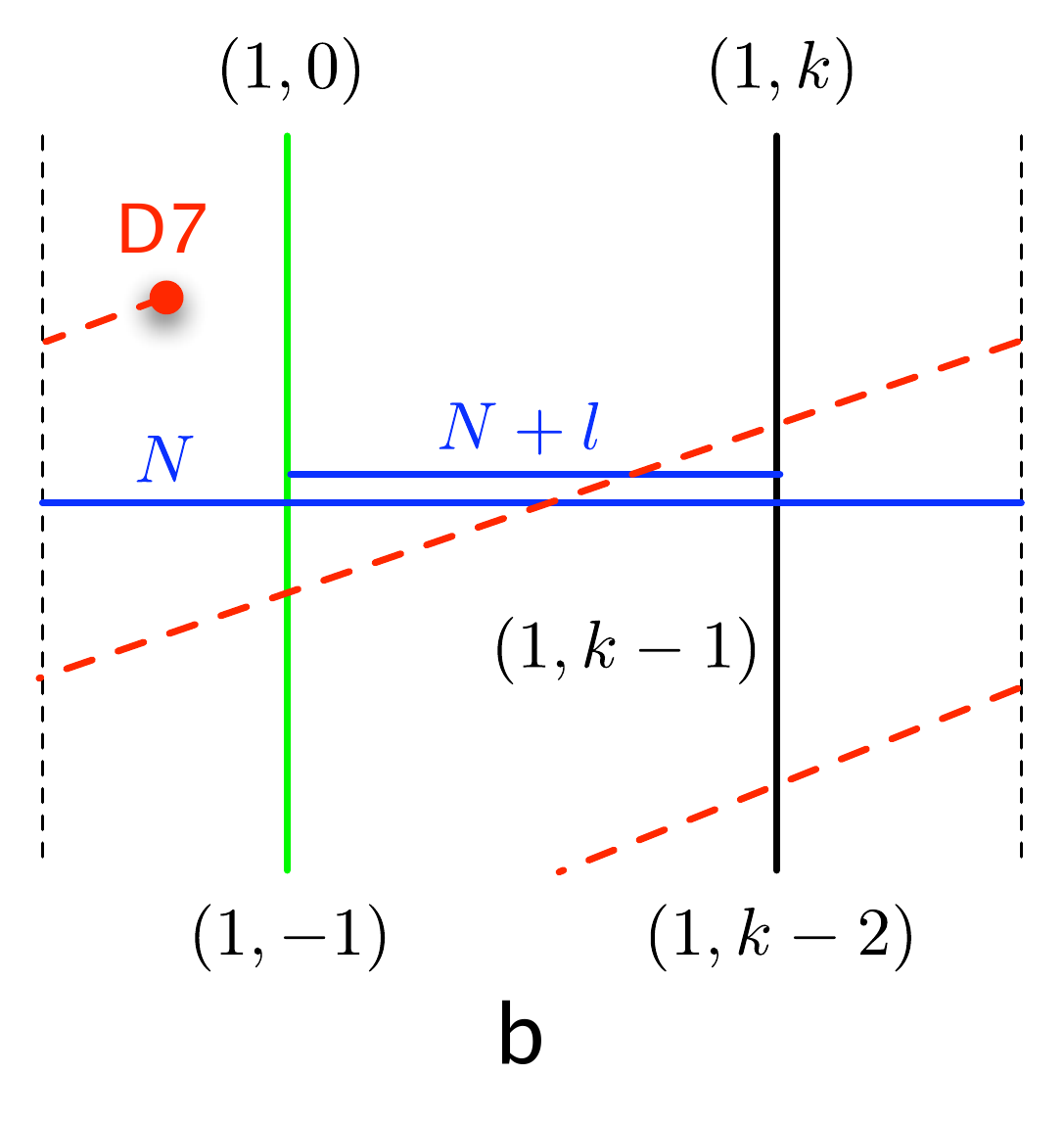,scale=0.4}
\caption{{\bf Different cuts for the same theory: $U(N+l)_k\times U(N)_{-k+1}$.}}
\label{other_cut}
\end{center}
\end{figure}

As we mentioned above, we are free to choose the direction of the cut from the D7-brane.
If the cut intersects either 5-brane, it transforms the 5-brane by its $SL(2,\mathbb{Z})$ monodromy,
which in this case just adds (or subtracts) one unit of D5-brane charge.
%
%
A generic choice for the cut will intersect the D3-branes once and the two 5-branes multiple times
(Fig.~\ref{other_cut}).
Each time it crosses a 5-brane it reduces its RR (D5) charge by one unit.
This could further shift both CS levels if there is an odd number of 5-brane crossings above the D3-branes.
However whenever this happens, the intersection of the cut and the D3-branes necessarily moves to 
the other side of the circle, and the net effect preserves both CS levels. 
For example in Fig.~\ref{other_cut}b we read off the field theory as
$U(N+l)_{(k-2)-(-1)+1}\times U(N)_{-1-(k-2)} = U(N+l)_k\times U(N)_{-k+1}$, which is the same 
as that of the original choice for the cut.

\begin{figure}[htbp]
\begin{center}
\epsfig{file=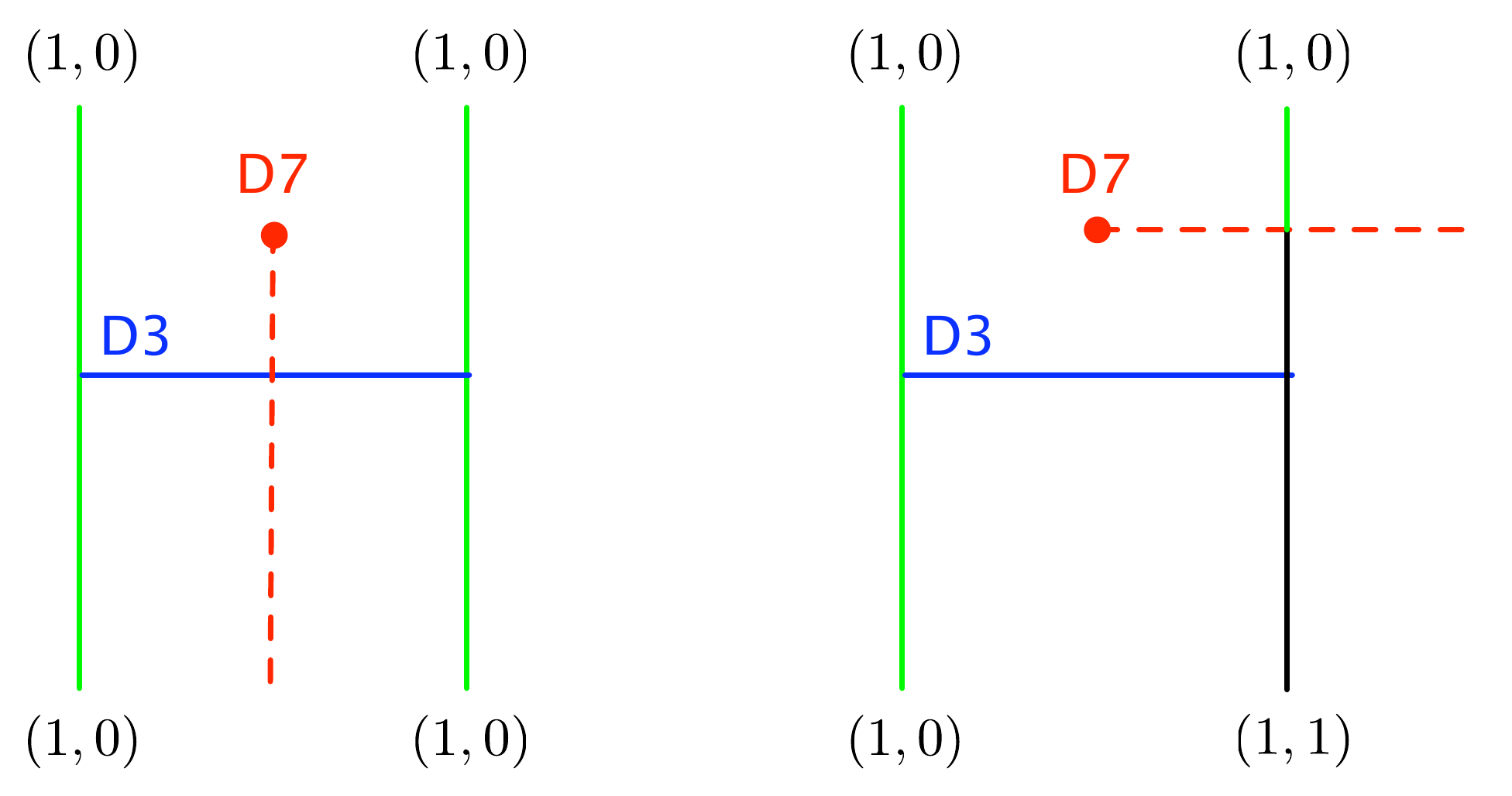,scale=0.4}
\caption{{\bf Alternative explanation of the CS term.}}
\label{CS_term_1}
\end{center}
\end{figure}

As an aside, note that we can use this construction to give an alternative derivation of
the original CS term on D3-branes suspended between 5-branes.
In \cite{KOO} the CS term was introduced in order to cancel a surface term in the 
variation with respect to the D3-brane worldvolume gauge field coming from the
boundary condition on a $(p,q)$5-brane.
In \cite{BHKK} this term was derived for the NS5-brane-$(1,k)$5-brane configuration, 
though somewhat indirectly, 
using two NS5-branes and $k$ D5-branes, by allowing the D5-branes to split
along one of the NS5-branes. In the 3d effective gauge theory this corresponds
to real mass terms for the fermions in the fundamental representation, 
which, when integrated out, lead to a level $k$ CS term.
Now we can derive the CS term more directly using string theory.
Start with a D3-brane suspended between a pair of parallel NS5-branes,
and add the D7-brane between the two NS5-branes (Fig.~\ref{CS_term_1}).
A vertical cut intersecting the D3-brane leads to a CS term as in (\ref{RR_CS}),
whereas a horizontal cut instead intersects one of the NS5-branes and transforms it into
a $(1,1)$5-brane. Therefore the 3d effective gauge theory on D3-branes
between an NS5-brane and a $(1,1)$5-brane contains a level 1 CS term.
This is easily generalized to an NS5-brane-$(1,k)$5-brane configuration.

\section{RR charges and field theory parameters}

The Type IIB brane configurations described in the previous section give
a four parameter family of UV gauge theories that can be characterized by 
their gauge groups and CS levels as $U(N+l)_k\times U(N)_{-k+q}$.
In the dual Type IIA picture these correspond to backgrounds with 
four kinds of RR flux: $F_6, F_4, F_2$ and $F_0$.
The question we would like to address in this section is how precisely 
the fluxes are related to the field theory parameters.
But first we should clarify what we mean by the RR fluxes, as there is more than one definition. 


As stressed in \cite{GT,AHHO}, 
the appropriate fluxes to compare with the field theory parameters are the ``Page charges".
These are defined by the modified field strengths:
\be
\label{Page_def}
\hat{F} = \tilde{F}\wedge e^{-B_2} \,,
\ee
where $\tilde{F}$ are the gauge-invariant fields of massive Type IIA supergravity, given by
\begin{equation}
\begin{array}{ll}
\tilde{F}_0 = F_0 & \\[5pt]
\tilde{F}_2 = dC_1 + F_0 B_2 & \;\;\;\; \tilde{F}_8 = *\tilde{F}_2 \\[5pt]
\tilde{F}_4 = dC_3 + C_1\wedge H_3 + {1\over 2} F_0 B_2\wedge B_2 & \;\;\;\; \tilde{F}_6 = *\tilde{F}_4 \,.
\end{array}
\end{equation}
The gauge invariant fields define the gauge invariant ``Maxwell charges".
However since $d\tilde{F}\neq 0$ generically, Maxwell charge is generically not quantized
and not conserved.
It is therefore not the appropriate quantity to compare with the integer field theory parameters.
The modified field strengths on the other hand satisfy $d\hat{F}=0$, 
so Page charge is conserved and quantized.
However it is not gauge invariant under the B-field gauge transformation $B_2\rightarrow B_2 + d\lambda$.
The Page charges in our case are given by fluxes on the cycles of $\mathbb{C}P^3$
\be
Q_{2n}^P = \int_{\mathbb{C}P^{4-n}} \hat{F}_{8-2n} \,,
\ee
for $n=1,2,3$ and 4, corresponding respectively to D2-branes, D4-branes on $\mathbb{C}P^1$,
D6-branes on $\mathbb{C}P^2$, and D8-branes on $\mathbb{C}P^3$.
These charges are well-defined up to large gauge transformations
that shift the B-field flux $b=\int_{\mathbb{C}P^1} B_2$ by an integer.
In particular, under $b\rightarrow b \pm 1$ the Page charges transform as
\begin{equation}
\begin{array}{ll}
\delta Q_2^P = \mp Q_4^P + {1\over 2} Q_6^P \mp {1\over 6} Q_8^P & \;\;\;\;\;\; 
\delta Q_6^{P} = \mp Q_8^P \\[5pt]
\delta Q_4^{P} = \mp Q_6^P + {1\over 2} Q_8^P & \;\;\;\;\;\; \delta Q_8^{P} = 0 \,.
\end{array}
\label{Page_trans}
\end{equation}


The transformation of the Page charges is related to the transformation that the field theory 
parameters undergo when one of the 5-branes, say the NS5-brane, is moved around the 
circle \cite{AHHO}.\footnote{To be precise,
the motion of the NS-brane once around the circle is T-dual in Type IIA to a continuous change of the
boundary value of the B field, such that $b_\infty\rightarrow b_\infty \pm 1$.
One can then interpret the change in the UV field theory parameters as the change 
in the Page charges under a gauge transformation $b \rightarrow b \mp 1$, that brings $b_\infty$
back to the range $[0,1]$.}
In supersymmetric situations,
continuous brane motions lead to different UV gauge theories which have the same IR dynamics,
and the two theories are said to be ``dual" or ``IR equivalent". 
An example of this in four dimensions is Seiberg duality, which can be derived
using a Type IIA brane configuration \cite{Elitzur}.
Similar relations can be obtained for the three dimensional ${\cal N}=3$ gauge theories by considering the
motion of the NS5-brane in Fig.~\ref{ABJ_configuration}.
In particular moving once around the circle to the right generates the ``parity duality"
between the theories with $U(N+l)_k\times U(N)_{-k}$ and $U(N)_k\times U(N+k-l)_{-k}$ \cite{ABJ}.
Other motions lead to more IR equivalences, and in particular imply that some
theories with $l>k$ flow in the IR to the ${\cal N}=6$ CFT's with $l\leq k$
\cite{AHHO,Evslin}.

\begin{figure}[htbp]
\begin{center}
\epsfig{file=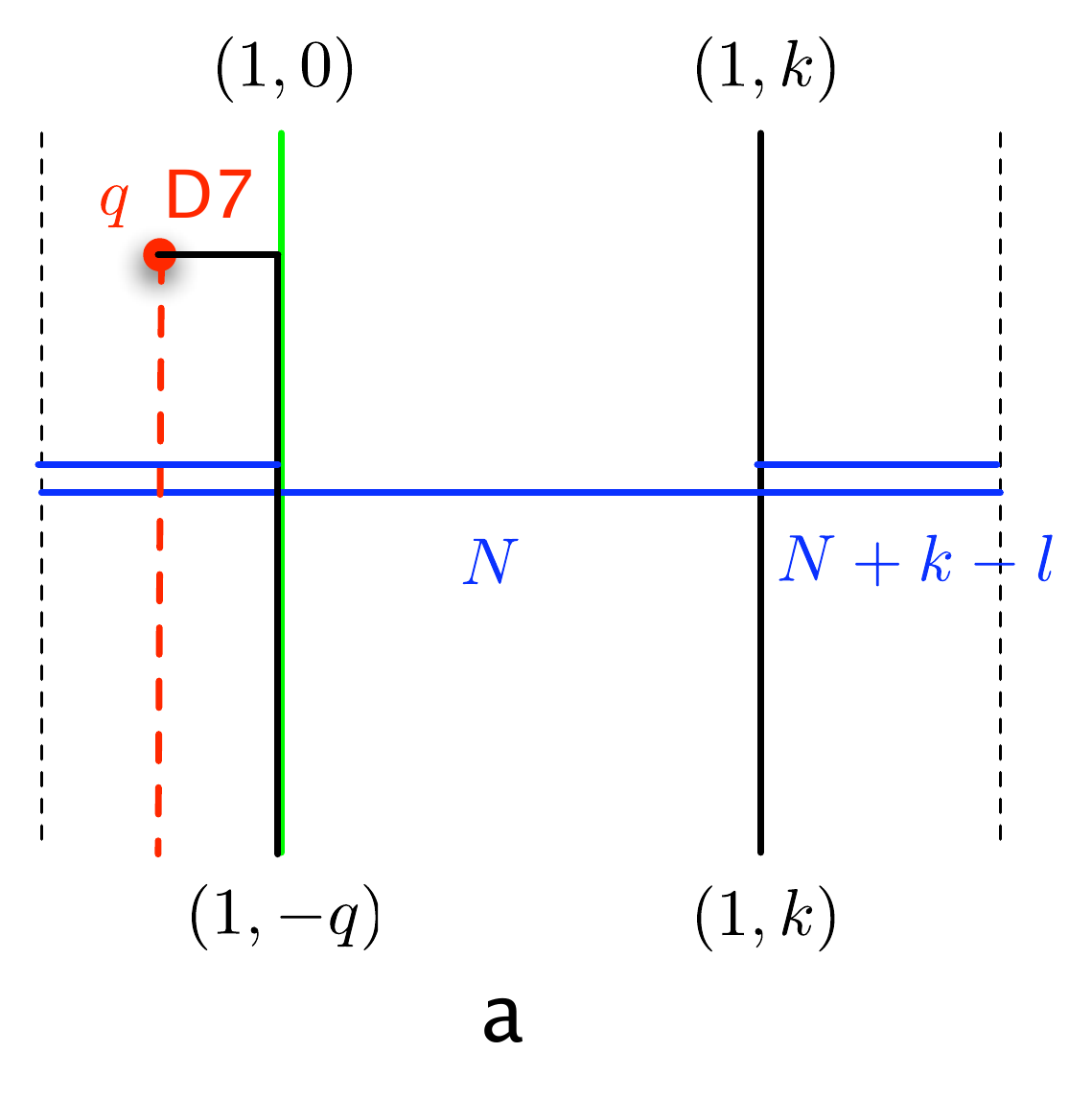,scale=0.4}
\hskip 1cm
\epsfig{file=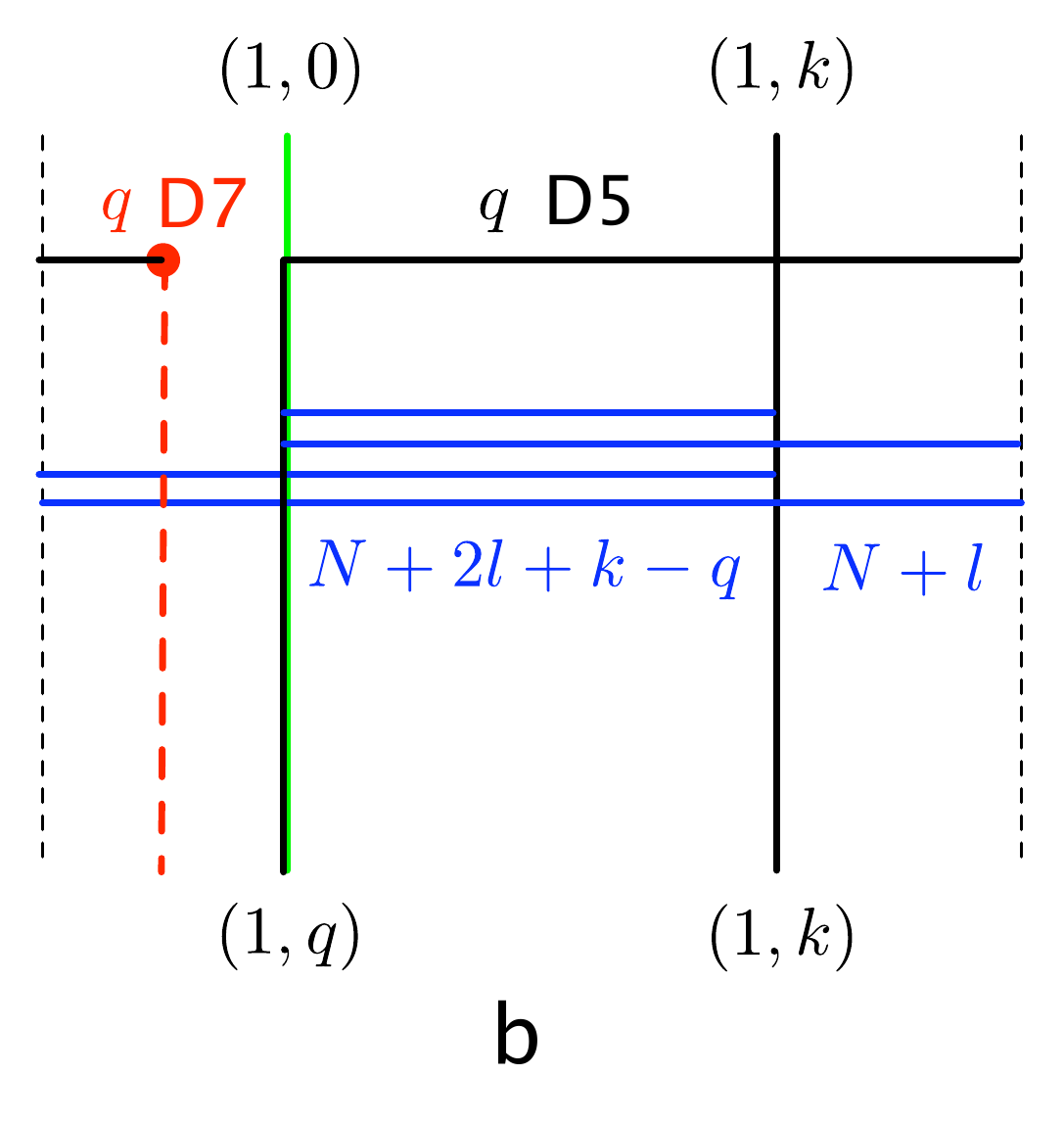,scale=0.4}
\caption{{\bf Moving the NS-brane around the circle to the right and to the left.}}
\label{motion}
\end{center}
\end{figure}

In generalizing to the configurations with D7-branes
we have to take into account two brane creation effects:
D3-branes are created when the two 5-branes cross, and D5-branes are created
when the NS5-brane crosses the D7-branes.
The latter changes the RR charge of the 5-brane by one unit for each D7-brane crossing.
Starting with the configuration for $U(N+l)_k\times U(N)_{-k+q}$ (Fig.~\ref{D7_deformation}a)
and moving the NS5-brane around the circle to the right changes the theory 
to $U(N)_{k+q} \times U(N-l+k)_{-k}$ (Fig.\ref{motion}a), in other words the field theory parameters
transform as
\be 
\label{rightward}
N\rightarrow N - l \,,\; l\rightarrow l-k \,,\; k\rightarrow k+q \,.
\ee
Moving the NS5-brane once around the circle to the left gives $U(N+2l+k-q)_{k-q} \times U(N+l)_{-k+2q}$
(Fig.~\ref{motion}b), {\em i.e.}
\be 
\label{leftward}
N\rightarrow N + l \,,\; l\rightarrow l+k \,,\; k\rightarrow k-q \,.
\ee
Note that in this case the NS5-brane becomes a $(1,q)$5-brane before it crosses the $(1,k)$5-brane,
leading to the creation of $(k-q)$ D3-branes.
In the absence of supersymmetry we cannot reliably conclude that the theories related 
by such motions are dual or IR equivalent.
However we can still use the relation to the transformation of the Page charges to determine
how the charges are related to the field theory parameters.

Consider the most general linear relation:
\be
\label{Page_general}
Q_2^P &=& N + \alpha_l l + \alpha_k k + \alpha_q q \nonumber \\
Q_4^P &=& l + \beta_k k + \beta_q q \nonumber \\
Q_6^P &=& k + \gamma_q q \\
Q_8^P &=& q \nonumber \,,
\ee
where the coefficients $\alpha_{l,k,q},\beta_{k,q},\gamma_q$ correspond to 
the possible contributions to the charge of a brane coming from higher dimensional branes.
Comparing the Page charge transformations with the transformations of the field theory parameters
(where we identify $b\rightarrow b \pm 1$ with the transformation under rightward and leftward
motion, respectively), leads to constraints on these coefficients:
\be
\label{constraints}
\alpha_l - \beta_k -{1\over 2} &=& 0 \nonumber \\
\beta_k - \gamma_q + {1\over 2} &=& 0  \\
\alpha_l + \alpha_k - \beta_q - {1\over 2}\gamma_q - {1\over 6} &=& 0 \nonumber \,.
\ee
Clearly the naive relation with all vanishing coefficients is not possible.
However these constraints are not sufficient to determine the coefficients.

To obtain additional constraints let us examine more closely the sources for these charges,
namely D-branes wrapping cycles of $\mathbb{C}P^3$ and forming radial domain walls
in $AdS_4$ \cite{BH}.
Each brane produces a jump in one, or more, of the field theory parameters.
On the other hand this D-brane can carry charges of lower-dimensional D-branes as well,
due to the worldvolume CS and curvature terms.
We can then compare the two effects to obtain conditions on the coefficients in (\ref{Page_general}).
The D$(2m)$-brane Page charge on a D$(2n)$-brane domain wall is given by
\be
Q^P_{2m,2n} = {1\over (2\pi)^{2(n-m)}} \int_{\mathbb{C}P^{n-m}}
\left[e ^{2\pi F} \wedge 
\sqrt{\hat{\cal A}(4\pi^2 R_T)\over \hat{\cal A}(4\pi^2 R_N)} \right]_{2(n-m)}\,,
\ee
where $\hat{\cal A}$ denotes the ``A-roof" (or Dirac) genus, and $R_T, R_N$ denote the curvatures
of the tangent and normal bundles to the D-brane worldvolume, respectively.
For our purpose only the first non-trivial term in $\hat{\cal A}$ is relevant:
\be
\hat{\cal A} = 1 - {1\over 24}p_1 + \cdots \,.
\ee
Note that by its definition (\ref{Page_def}), the Page charge is independent of the B-field.

The curvature contribution is fixed by the $\mathbb{C}P^n$ cycle that the D$(2n)$-brane wraps.
In particular it is trivial for the D2-brane and the D4-brane on $\mathbb{C}P^1$.
For the D6-brane on $\mathbb{C}P^2$ the curvature term contributes to the D2-brane charge an amount
${1\over 48}(p_1(N(\mathbb{C}P^2))-p_1(T(\mathbb{C}P^2)) = -{1\over 24}$,
and for the D8-brane on $\mathbb{C}P^3$ it contributes to the D4-brane charge an amount $-{1\over 48}p_1(T(\mathbb{C}P^3))=-{1\over 12}$.

The contribution of the worldvolume field strength $F$ is not fixed a-priori, and we can consider different,
but quantized, amounts of flux on the $\mathbb{C}P^1$ in the D-brane domain wall.
The usual Dirac quantization condition requires $\int_{\mathbb{C}P^1}(F/2\pi)\in \mathbb{Z}$.
However for the D6-brane on $\mathbb{C}P^2$ this is shifted to
$\int_{\mathbb{C}P^1}(F/2\pi)\in \mathbb{Z} + {1\over 2}$ due to an anomaly 
associated with non-spin manifolds \cite{FW}.
Let us take the minimal amount of flux on the domain wall D-branes.
For the D2, D4 and D8 this is 0, and for the D6 this is $\pm 1/2$.
We will assume that this choice corresponds to the situation where only one of the parameters
jumps for each D-brane, namely $N\rightarrow N+1$ for the D2-brane, $l\rightarrow l+1$
for the D4-brane, $k\rightarrow k+1$ for the D6-brane, and $q\rightarrow q+1$ for the D8-brane.
This assumption will be justified by the consistency of the result with the constraints in (\ref{constraints}).

Adding the curvature and flux contributions gives the lower brane charges shown in table~1.
This fixes all the coefficients in (\ref{Page_general}) up to one sign:
\be
\label{coefficients}
\alpha_l = \alpha_q = \gamma_q = 0 \; , \; 
\beta_k = \pm {1\over 2} \; , \;
\alpha_k = -\beta_q = {1\over 12} \,,
\ee
and it is easy to see that these satisfy (\ref{constraints}) if $\beta_k=-1/2$.
We conclude that the charge/parameter relation is given by
\be
\label{Page_charges}
Q_2^P & = & N + {k\over 12} \nonumber \\
Q_4^P &=& l - {k\over 2} - {q\over 12} \nonumber \\
Q_6^P &=& k \\
Q_8^P &=& q \nonumber \,.
\ee
This agrees with what was found in \cite{AHHO} for $q=0$.

\begin{table}
\label{lower_brane_charges}
\begin{center}
\begin{tabular}{|c|c|c|}
\hline
Domain wall & parameter change & lower branes \\
\hline
D2 & $N\rightarrow N+1$ & none \\
D4 on $\mathbb{C}P^1$ & $l\rightarrow l+1$ & none \\
D6 on $\mathbb{C}P^2$ & $k\rightarrow k+1$ & $\pm {1\over 2}$ D4 + ${1\over 12}$ D2\\
D8 on $\mathbb{C}P^3$ & $q\rightarrow q+1$ & $-{1\over 12}$ D4\\
\hline
\end{tabular}
\caption{Domain wall branes and lower brane charges.}
\end{center}
\end{table}

\section{Conclusions}

We presented a Type IIB brane realization of a non-supersymmetric three dimensional 
Yang-Mills-Chern-Simons gauge theory
with $U(N_1)_{k_1}\times U(N_2)_{k_2}$, matter in the bi-fundamental representation,
and a global $SO(3)$ symmetry.
The configuration is a deformation of the configuration that describes the ${\cal N}=3$
gauge theory with $U(N_1)_{k}\times U(N_2)_{-k}$. It is known that the ${\cal N}=3$ 
theory flows in the IR to the ${\cal N}=6$ 
superconformal Chern-Simons-matter theory, which is dual to the $AdS_4\times\mathbb{C}P^3$
solution of Type IIA supergravity.
One would like to conclude that the non-supersymmetric gauge theory likewise flows in the IR
to the $SO(6)$ invariant non-supersymmetric conformal CSM theory, which was conjectured
in \cite{GT} to be dual to an $AdS_4\times\mathbb{C}P^3$ solution of massive Type IIA supergravity.
This would require the CFT to be attractive in the space of $SO(3)$ invariant theories.
In the ${\cal N}=6$ case this is guaranteed by supersymmetry, 
but in the ${\cal N}=0$ case it is not.\footnote{We thank Ofer Aharony for discussions on this point.}
On the other hand, the relation that we exhibited between the Type IIB brane configuration and the 
massive Type IIA background suggests that this is true, and
provides further evidence for the conjecture of \cite{GT} in the ${\cal N}=0$ case.
It is of course desirable to find more evidence.
It would also be interesting to better understand the ${\cal N}=1$ case,
possibly relating it to the ${\cal N}=1$ massless Type IIA background of \cite{Ooguri},
and to find the complete ${\cal N}=2$ and 3 solutions.

\appendix

\section{Geometry of $\mathbb{C}P^3$}

The complex projective space $\mathbb{C}P^3$ is defined as
\be
\mathbb{C}P^3 = {\mathbb{C}^4\over z_i \sim \lambda z_i}\; ,\; i=1,\ldots ,4 \;,\;\lambda\in\mathbb{C} \,.
\ee
We can fix $\sum_i|z_i|^2=1$ and represent $\mathbb{C}P^3$
as the quotient of $S^7$ by a $U(1)$ action
\be
\mathbb{C}P^3 = {S^7\over z_i \sim e^{i\varphi} z_i}\,.
\ee
The $S^7$ can be parametrized as follows:
\be
z_1 &=& \cos\xi \cos(\theta_1/2) e^{i(\psi_1+\phi_1)/2} \\
z_2 &=& \cos\xi \sin(\theta_1/2) e^{i(\psi_1-\phi_1)/2} \\
z_3 &=& \sin\xi \cos(\theta_2/2) e^{i(\psi_2+\phi_2)/2} \\
z_4 &=& \sin\xi \sin(\theta_2/2) e^{i(\psi_2-\phi_2)/2} \,,
\ee
where $0\le\xi\le\pi/2$, $0\le\theta_i\leq\pi$, $0\le\phi_i<2\pi$, and $0\le\psi_i<4\pi$.
The round metric on $S^7$ is then given by
\be
ds^2_{S^7} = d\xi^2 
&+& {\cos^2\xi\over 4}
\left[\left(d\psi_1 + \cos\theta_1 d\phi_1\right)^2 + d\theta_1^2 + \sin^2\theta_1 d\phi_1^2 \right]\nonumber \\
&+& {\sin^2\xi\over 4}\left[\left(d\psi_2 + \cos\theta_2 d\phi_2\right)^2 + d\theta_2^2 + \sin^2\theta_2 d\phi_2^2 \right]\,.
\ee
In this parameterization the $S^7$ is represented as $S^3\times S^3$ fibered over an interval
(parameterized by $\xi$), where each $S^3$
is represented as an $S^1$ fibered over an $S^2$.
We can rewrite this as an $S^1$ fibered over $\mathbb{C}P^3$ by defining
\be
\psi_1 = 2\varphi + \psi \;,\; \psi_2 = 2\varphi - \psi \,,
\ee
where $0\leq \varphi <2\pi$ and $0\leq \psi < 2\pi$. Then
\be
ds^2_{S^7} = (d\varphi + \omega)^2 + ds^2_{\mathbb{C}P^3} \,,
\ee
where 
\be
\omega={1\over 2}(\cos^2\xi-\sin^2\xi)d\psi+{1\over 2}\cos^2\xi\cos\theta_1d\phi_1
+ {1\over 2}\sin^2\xi\cos\theta_2d\phi_2 \,,
\ee
and 
\be
ds_{\mathbb{C}P^3}^2&=&d\xi^2+\cos^2\xi\sin^2\xi\left(d\psi+{\cos\theta_1\over 2}d\phi_1-{\cos\theta_2\over 2}d\phi_2\right)^2\nonumber\\
&&+{1\over 4}\cos^2\xi\left(d\theta_1^2+\sin^2\theta_1d\phi_1^2\right)
+{1\over 4}\sin^2\xi\left(d\theta_2^2+\sin^2\theta_2d\phi_2^2\right) \,. \label{CP3}
\ee
In this parameterization $\mathbb{C}P^3$ is represented as a $T^{1,1}$ fibered over an interval,
where the $T^{1,1}$ is an $S^1$ (parameterized by $\psi$) fibered over $S^2\times S^2$.
$\mathbb{C}P^3$ has a 4-cycle $\mathbb{C}P^2$, corresponding to fixed $\theta_2$
and $\phi_2$:
\be
ds^2_{\mathbb{C}P^2} = d\xi^2 + \cos^2\xi \sin^2\xi \left(d\psi + {\cos\theta_1\over 2}d\phi_1\right)^2
+ {\cos^2\xi\over 4}\left(d\theta_1^2 + \sin^2\theta_1 d\phi_1^2\right),
\ee
and a 2-cycle $\mathbb{C}P^1$ corresponding to fixed $\theta_1, \theta_2, \phi_1, \phi_2$:
\be
ds^2_{\mathbb{C}P^1} = d\xi^2 + {1\over 4}\sin^2(2\xi) d\psi^2 
= {1\over 4}\left(d(2\xi)^2 + \sin^2(2\xi)d\psi^2\right)\,.
\ee

\section*{Acknowledgment}

We thank the Aspen Center for Physics, where some of the work was done.
We would like to thank Ofer Aharony for his insightful comments on the manuscript,
and Daniel Jafferis, Igor Klebanov and Juan Maldacena for useful discussions.
The work of OB and GL was supported in part by the Israel Science Foundation under grant no.~568/05.

\end{document}